\documentclass[12pt]{iopart}

\usepackage{mathtext}
\usepackage{graphicx,cite}
\usepackage{url}
\usepackage{epstopdf}

\usepackage{color}

\newcommand{\vvec}[1]{\mathbf{#1}}
\newcommand{\mathbb}[1]{\mathbf{#1}}

\begin{document}

\title[Universal shape characteristics]{Universal shape characteristics for the mesoscopic star-shaped polymer \textit{via} dissipative particle dynamics simulations}

\author{O.~Kalyuzhnyi$^{1,4}$, J.M.~Ilnytskyi$^{1,4}$, Yu.Holovatch$^{1,4}$}
\author{C. von Ferber$^{2,3,4}$}

\address{$^{1}$Institute for Condensed Matter Physics, National Academy of Sciences of Ukraine, UA--79011 Lviv, Ukraine}
\address{$^{2}$Applied Mathematics Research Centre, Coventry University, Coventry, CV1 5FB, United Kingdom}
\address{$^{3}$Heinrich-Heine Universit\"at D\"usseldorf, D-40225 D\"usseldorf, Germany}
\address{$^{4}$${\mathbb L}^4$ Collaboration \& Doctoral College for the Statistical
Physics of Complex Systems, Leipzig-Lorraine-Lviv-Coventry,
D-04009  Leipzig, Germany}

\begin{abstract}
In this paper we study the shape characteristics of star-like
polymers in various solvent quality using a mesoscopic level of
modeling. The dissipative particle dynamics simulations are
performed for the homogeneous and four different heterogeneous star
polymers with the same molecular weight. We analyse the gyration
radius and asphericity at the bad, good and $\theta$-solvent
regimes. Detailed explanation based on interplay between enthalpic and
entropic contributions to the free energy and analyses on of the asphericity of individual
branches are provided to explain the increase of the apsphericity in $\theta$-solvent regime.
\end{abstract}

%
\vspace{2pc} \noindent{\it Keywords}: star-like
polymer, $\theta$-solvent, dissipative particle dynamics
%

\submitto{\JPCM}
%

\date{\today}

\maketitle

\section{\label{I}Introduction}
Star polymers represent the simplest case of branched polymers,
consisting of $f$ linear chains connected to the central core.
The central core can be an atom, or a molecule or even a macromolecule.
The first synthesis of star polymers was made by Shaefgen and Flory
\cite{Flory1948} in 1948. More than ten years later Morton \emph{et
al}. \cite{Morton} have synthesized the 3- and 4-branch polystyrene
stars. Since then the methods of synthesis have been drastically
improved. Currently, there are several general methods ti synthesis
star polymers: using multifunctional linking agents or
multifunctional initiators and via difunctional monomers
\cite{Mishra,Hadji2001}. There are several important polymerization
techniques, i.e. cationic, anionic, controlled radical, ring
opening, group transfer, step-growth polycondensation and their
combinations (see Ref. \cite{Handji2012} and references therein).
Substantial interest in theoretical and experimental studies of star
polymers arises due to their important applications.
by their practical applications importance. Star polymers may
have a properties much different from those of the linear chain
polymers. They have more compact size and, therefore, higher segment
density compared to their linear counterparts with the same
molecular weight. This feature affects substantially the properties
of the star polymer containing systems \cite{Mishra,Hadji2001}. Their
bulk viscosity in concentrated as well as in dilute solution, can be lower than for a
linear polymers of similar molecular weight. Besides that star polymers can
self-assemble in more types of microstructures, which can be
promoted in solutions due to the presence of the functional groups
in their branches, or by using a selective solvent in the case of
star-block or miktoarm star copolymers, leading to the formation of
micellar structures \cite{Khanna2010}.

In terms of molecular topology, star polymers represent an
intermediate system between linear chain polymers and colloidal
particles such as polystyrene and silica spheres. This feature has
been discussed in a number of papers, in which the structure
\cite{Gast,Grest,Likos98,Witten} and dynamics \cite{Sikorski93,
Factor90, Vlassopoulos99,
 Grest1989, Su91, Su92, STRATING1994, Sikorski99, Ganazzoli95}
of the star polymers have been studied. The properties of the stars
with a small number of branches are similar to those of the
 linear chain polymers. In particular, their
average configuration is characterized by a large aspherisity
\cite{Solc1971a,Solc1971b}. The star structures with larger number
of the branches have much lower aspherisity and in the
limit of large $f$ they can be seen as rigid spherical particles
\cite{Zifferer1999a}.
 Star polymers have substantially
higher shear stability \cite{Xue2005} and they are widely used as
viscosity index modifier in the multigrade lubricating oils
\cite{Lohse02, Schober2008}. They are also used for
manufactoring the thermoplastic elastomers, which at room
temperature have the properties similar to those of cross-linked
elastomers (such as vulcanized rubber). However, with the
temperature increase, they became soft and can flow, which is a very
useful property for their processing \cite{Knoll1998}. In addition,
thermoplastic elastomers, in contrast to rubber, can be reused. For
producing thermoplastic elastomers the mixture of linear and
miktoarm star triblock copolymers is often used \cite{Knoll1998}.
Star polymer systems are used in coating materials, as binders for
toners for copying machines \cite{Grest}, resins for
electrophotographic photoreceptors \cite{Maness1997} and in a number
of pharmaceutical and medical applications \cite{Likos,Grest}. Star
polymers and starburst dendrimers also have important applications
in semiconductor devices \cite{Hecht1999}, in biofunctionalized
patterning of the surfaces \cite{Heyes2007}, and in controlled
release drug delivery \cite{PriyaJames2014}. Reviews of the
synthesis, properties and applications of star polymers are given in
Ref.~\cite{Handji2012}.

Due to the substantial advances in the synthesis of star polymers,
as well as their numerous applications, they have attracted
considerable interest both theoretical and experimental methods
\cite{Mishra,Hadji2001,Grest,Handji2012,Khanna2010}. Theoretical
studies of the polymer stars were carried out using
renormalization group \cite{Douglas90, Shiwa, Merkle, Zhu90, Vlahos,
Dupla, Binder91, Binder88, Miyake83, Miyake84, Vlahos87, vonFerber1997} (see also
Ref. \cite{Holovatch2002} and references therein) and the field
theoretical \cite{Chujo, Okamoto, Knoll} approaches, extrapolation
of exact enumerations \cite{Lipson, Wilkinson, Barrett, Colby}, free
energy minimization method \cite{Ganazzoli90, Ganazzoli91,
Ganazzoli92}, mean field \cite{Boothroyd90, Boothroyd91, Irvine} and
scaling theories \cite{Ohno91, Ohno9, Halperin, Raphael, Daoud,
Birshtein, Duplantier, Duplantier86, Duplantier8}, as well as the
density functional approach \cite{Groh}. In
general, the properties of polymers have been studied intensively in
the past using computer simulations methods, such as molecular
dynamics and Monte-Carlo methods \cite{Rey92, Binder95, Frenkel02,
Kotelyanskii, Galiatsatos}. These methods were also used to study
the properties of the star polymers and a large number of
corresponding studies have been published during the last decade
(see Ref. \cite{Lue} and references there in).

   One of important properties of star polymers that
affect their technological application is their shape.
Conformational and shape properties of the star polymers, which
include mean square radius of gyration and hydrodynamic radii, were
studied experimentaly \cite{Allgaier, Feng, Roovers72, Roovers74,
Zhou}. The universal shape properties of the linear chain polymers
in a form of a self-avoiding walk (SAW) were also studied recently
via lattice modeling \cite{Aronovitz1986, Cannon,
Zifferer1999a,Zifferer1999b, Jagodzinski1992, Bishop1988,
Benhamou1985, Diehl1989, Blavatska2011}. It was demonstrated that
certain shape properties of polymer chains, similar to the scaling exponents,
are universal and depend solely on dimensionality of the space $d$.
This type of the approach appears to be very efficient and allows
one to achieve very good configuration statistics by means of a
relatively low computational cost. However, the lattice models are not
able to easily account for the effects of chain stiffness, chain
composition, the quality of a solvent, etc. On the other hand,
atomistic first principal simulations \cite{Nardai2009}, which can
include all these features, require substantial amount of the
computer time. Perhaps, the most adequate computer simulation method
to be used, which is able to compromise between chemical versatility
and computational efficiency, is the dissipative particle dynamics
(DPD) method. Applications of this method to the case of star
polymers are already present
 \cite{Qian, Vliet, Ilnytskyi2007, Xia, Chou, Sheng, Espaol2017}. In particular, in the
paper of Nardai and Zifferer \cite{Nardai2009}, dilute solutions of
linear star polymers have been studied using DPD simulations. On the
other hand, in our previous paper \cite{Kalyuzhnyi2016} we applied the
DPD method to investigate the shape characteristics of a linear
polymer chain in a good solvent. We have shown that such shape
properties of the chain as asphericity, prolateness and some other
are universal. The paper may be regarded as an extension of our work
\cite{Kalyuzhnyi2016} on shape properties of linear polymer chains aiming at systematic
analysis of the combined impact of solvent quality and polymer topology of the
polymer shape properties. The aim of this study is to apply the same analysis
to the case of star polymers. The outline of the paper is as
follows. Sec.~\ref{II} contains detailed description of the model
star polymers and the properties of interest to be studied.
Sec.~\ref{III} contains the results and their interest. Conclusions
are given in Sec.~\ref{IV}.

\section{\label{II}The model, simulation approach and properties of interest}
As already mentioned in Sec.~\ref{I}, our study is based on the
mesoscopic DPD simulation technique. In this approach, the monomers,
as well as the solvent beads, represent respective groups of atoms. They
are soft spheres of equal size, repulsive to each other.Therefore, the DPD
technique enables one to consider both the effects of a star-like topology
as well as solvent quality via explicit account of mesoscopic polymer-solvent
and solvent-solvent interactions
\begin{figure}[!h]
\centerline{\includegraphics[width=0.24\textwidth,angle=270]{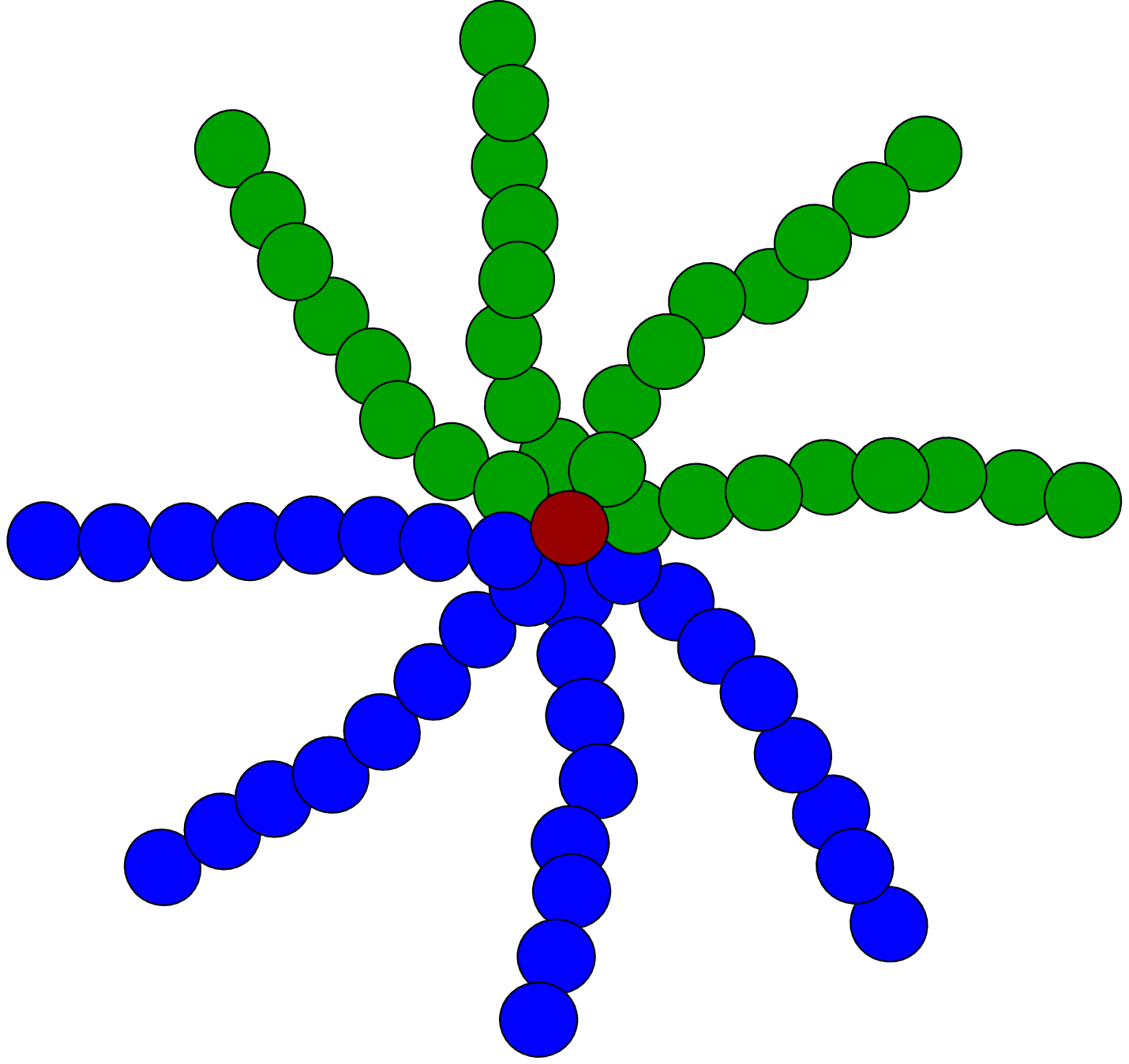}\hspace{3em}
\includegraphics[width=0.25\textwidth,angle=270]{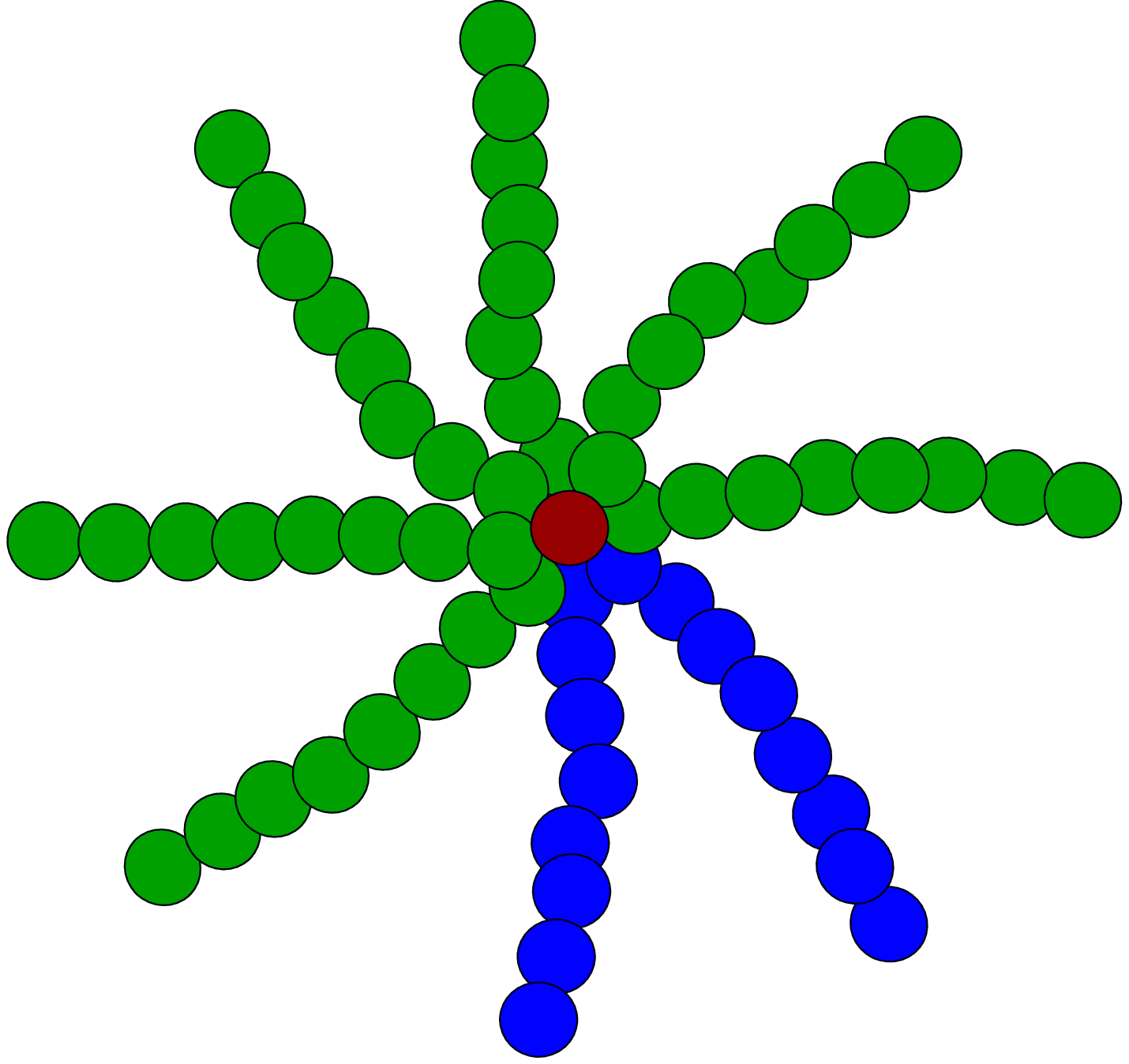}\hspace{3em}
\includegraphics[width=0.25\textwidth,angle=270]{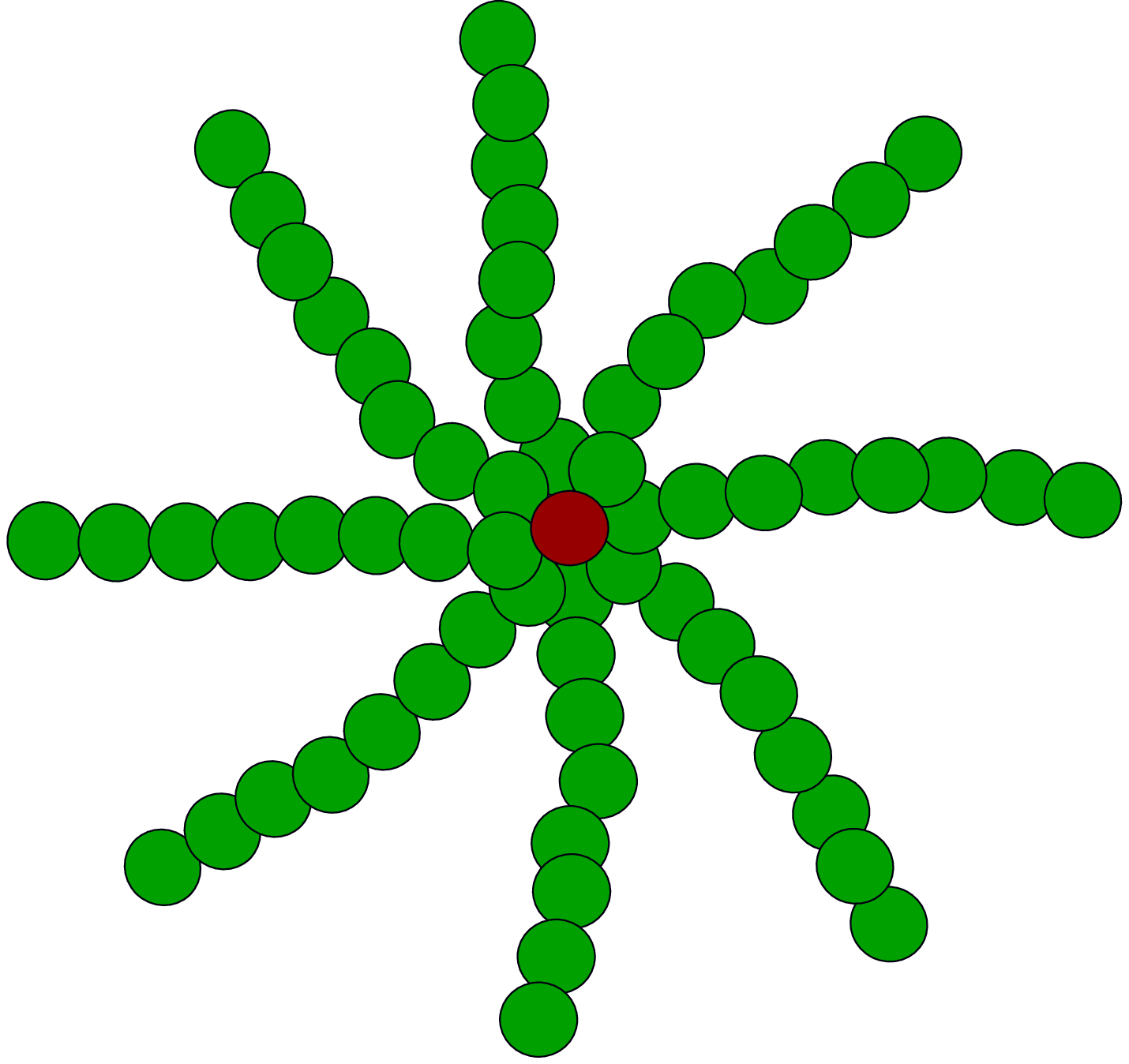}}
\centerline{4:0 \hspace{11em} 2:0 \hspace{11em} 0:0}
\end{figure}

\begin{figure}[!h]
\centerline{\includegraphics[width=0.25\textwidth,angle=270]{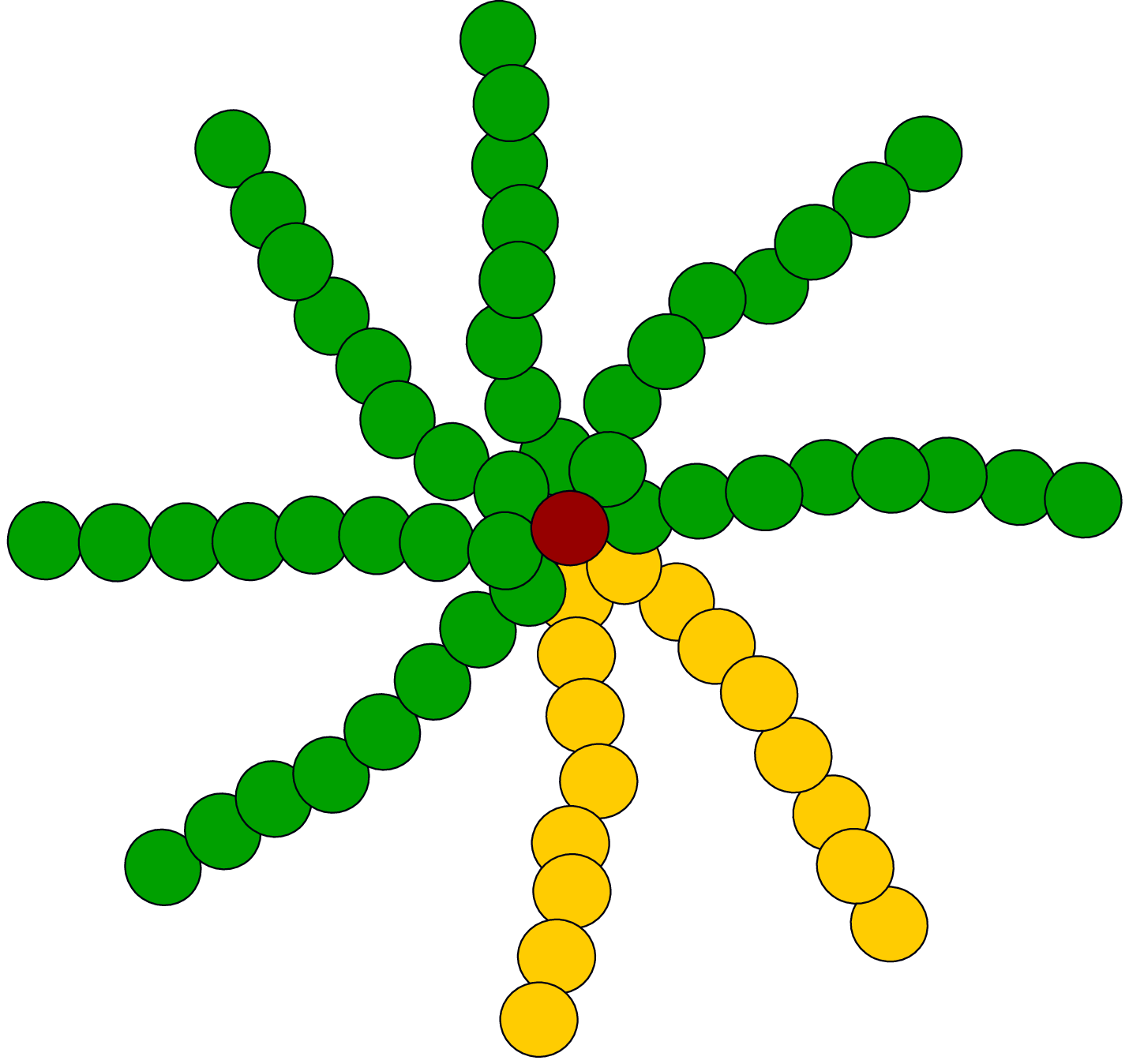}
\includegraphics[width=0.25\textwidth,angle=270]{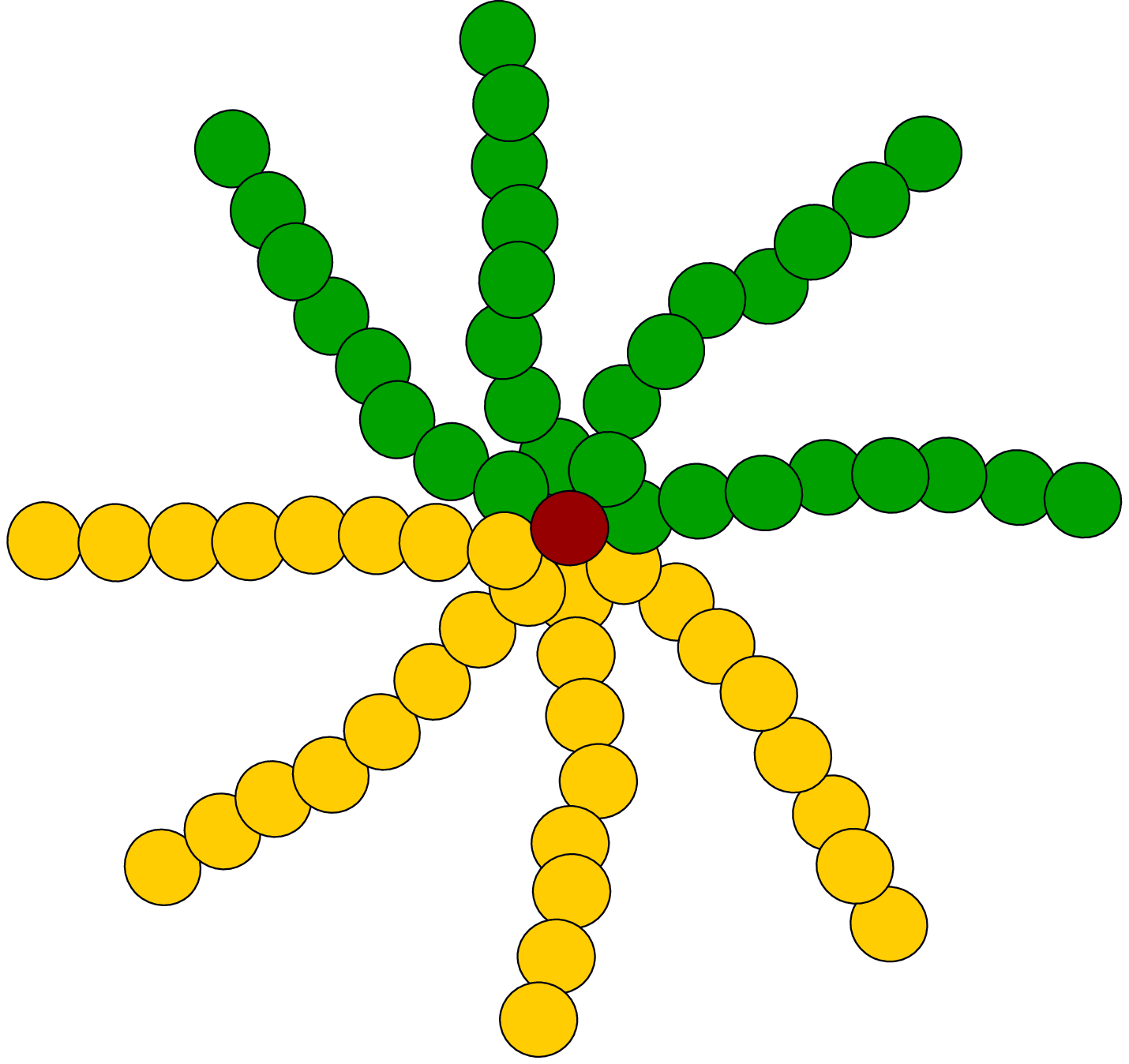}}
\centerline{0:2 \hspace{9em} 0:4} \caption{\label{Stars} Sketches of
polymer stars considered in this paper. Each star consists of
$f_A+f_B+f_C=8$ branches of equal length of $N_f=8$ monomers
attached to the central core (red sphere). $f_A$ and $f_B$ denotes
the number of solvophilic and solvophobic branches, respectively, $f_C$
stands for branches with variable solvophobicity. For the
homogeneous star ($f_A=0$, $f_B=0$, further denoted as 0:0), all
$f_C=8$ branches have the same solvophobicity, that
varies from the good solvent to bad solvent regime. For the heterogeneous stars
(4:0, 2:0, 0:2, 0:4), the blue $f_A$ branches do not change their
solvophobicity remaining always solvophilic while the yellow $f_B$
branches are always solvophobic, whereas solvophobicity of the green
$f_C$ branches varies.}
\end{figure}

We follow DPD method as formulated in Ref.~\cite{Groot1997}. The
following reduced quantities are used: the length will be measured
in units of the diameter of the soft bead, and the energy in units
of $k_{B}T$. Here, $k_{B}$ is the Boltzmann constant and $T$ is the
temperature. The monomers of each
branch are connected via harmonic springs, which results in the
force
\begin{equation}\label{FB}
  \vvec{F}^B_{ij} = -k\vvec{x}_{ij}\,,
\end{equation}
where $\vvec{x}_{ij}=\vvec{x}_i-\vvec{x}_j$, $\vvec{x}_i$ and
$\vvec{x}_j$ are the coordinates of $i$th and $j$th bead,
respectively, and $k$ is the spring constant. The total force acting
on the $i$th bead from the $j$th one can be denoted as
\begin{equation}
  \vvec{F}_{ij} = \vvec{F}^{\mathrm{C}}_{ij} + \vvec{F}^{\mathrm{D}}_{ij}
  + \vvec{F}^{\mathrm{R}}_{ij}\,,
\end{equation}
where $\vvec{F}^{\mathrm{C}}_{ij}$ is the conservative force
responsible for the repulsion between the beads,
$\vvec{F}^{\mathrm{D}}_{ij}$ is the dissipative force that defines
the friction between them and the random force
$\vvec{F}^{\mathrm{R}}_{ij}$  works in pair with a dissipative force
to thermostat the system. The expression for all these three
contribution are given below \cite{Groot1997}
\begin{equation}\label{FC}
  \vvec{F}^{\mathrm{C}}_{ij} =
     \left\{
     \begin{array}{ll}
        a(1-x_{ij})\displaystyle\frac{\vvec{x}_{ij}}{x_{ij}}, & x_{ij}<1,\\
        0,                       & x_{ij}\geq 1,
     \end{array}
     \right.
\end{equation}
\begin{equation}\label{FD}
  \vvec{F}^{\mathrm{D}}_{ij} = -\gamma
  w^{\mathrm{D}}(x_{ij})(\vvec{x}_{ij}\cdot\vvec{v}_{ij})\frac{\vvec{x}_{ij}}{x^2_{ij}},
\end{equation}
\begin{equation}\label{FR}
  \vvec{F}^{\mathrm{R}}_{ij} = \sigma
  w^{\mathrm{R}}(x_{ij})\theta_{ij}\Delta t^{-1/2}\frac{\vvec{x}_{ij}}{x_{ij}},
\end{equation}
where $x_{ij}=|\vvec{x}_{ij}|$,
$\vvec{v_{ij}}=\vvec{v}_{i}-\vvec{v}_{j}$, $\vvec{v}_{i}$ is the
velocity of the $i$th bead, $a$ is the amplitude for the conservative
repulsive force. The dissipative force has an amplitude $\gamma$ and
decays with distance according to the weight function
$w^{\mathrm{D}}(x_{ij})$. The amplitude for the random force is
$\sigma$ and the respective weight function is
$w^{\mathrm{R}}(x_{ij})$. $\theta_{ij}$ is the Gaussian random
variable. As was
shown by Espa\~{n}ol and Warren \cite{Espanol1995}, to satisfy the
detailed balance requirement, the amplitudes and weight
functions for the dissipative and random forces should be
interrelated: $\sigma^2=2\gamma$ and
$w^{\mathrm{D}}(x_{ij})=\large[w^{\mathrm{R}}(x_{ij})\large]^2$.
Here we use quadratically decaying form for the weight functions:
\begin{equation}
w^{\mathrm{D}}(x_{ij})=\large[w^{\mathrm{R}}(x_{ij})\large]^2
=\left\{
\begin{array}{ll}
(1-x_{ij})^2, & x_{ij} < 1,\\
0, & x_{ij} \geq 1.
\end{array}
\right.
\end{equation}
The reduced density of the system is defined as $\rho^{*} = (N +
N_{s})/V=3$ , where $N_s$ is the number of solvent particles and $V$
is system volume. The other parameters are chosen as follows:
$\gamma=6.75$, $\sigma=\sqrt{2\gamma}=3.67$

Parameter $a$ in the conservative force $\vvec{F}^{\mathrm{C}}_{ij}$
defines maximum repulsion between two beads which occurs at their
complete overlap, $x_{ij}=0$. We consider three types of branches:
solvophilic (beads of type $A$), solvophobic (beads of type $B$) and
with variable solvophobicity (beads of type $C$). Their respective
numbers are $f_A$, $f_B$ and $f_C$. Each star type is denoted as ($f_A:f_B$),
and $f_C$ is equal to $f_C=f-f_A-f_B$. The solvent beads are of type $S$.
The values of the repulsion parameter a are chosen according
to the bead types. Namely: $a_{AA}=a_{BB}=a_{CC}=a_{SS}=a_{AS}=25$  and
$a_{BS}=40$. The value of $a_{CS}$ is varied within the range from 25
(solvophilic branch) up to 40 (solvophobic branch). In what follows below, we have
chosen to characterize shape of star polymers in terms of their
radius of gyration and asphericity. The latter can be obtained
from the eigenvalues of the  gyration tensor
$\vvec{Q}$ defined as \cite{Solc1971a,Solc1971b}:

\begin{equation}\label{Q_def}
Q_{\alpha\beta} = \frac{1}{N}
\sum_{n=1}^{N}(x^{\alpha}_{n}-X^{\alpha})(x^{\beta}_{n}-X^{\beta})
\hspace{1em} \alpha,\beta=1,2,3,
\end{equation}
where $N$ is the number of monomers of a star polymer,
$x_n^{\alpha}$ stands for the set of the Cartesian coordinates of
$n$th  monomer: $\vvec{x}_n=(x_n^1,x_n^2,x_n^3)$, and
$X^{\alpha} = \frac{1}{N}\sum_{n=1}^N x_{n}^{\alpha}$ are the
coordinates of the center of mass for the star polymer. Its
eigenvectors define the axes of a local frame of a star polymer and
the mass distribution of the latter along each axis is given by the
respective eigenvalues $\lambda_i$, $i=1,2,3$. The
trace of $\vvec{Q}$ is an invariant with respect to rotations and is
equal to an instantaneous squared gyration radius of the star
polymer
\begin{equation}\label{Rg_def}
R_g^2 = \mathrm{Tr}\,\vvec{Q} =\sum_{i=1}^3\lambda_i = 3\bar{\lambda}
\end{equation}
Here, the arithmetic mean of three eigenvalues, $\bar{\lambda}$, is
introduced to simplify the forthcoming expressions. The asphericity $A$
\cite{Aronovitz1986,Zifferer1999a,Zifferer1999b,Blavatska2011} for the three-dimensional case is defined as
\begin{equation}\label{ASg_def}
A =
\frac{1}{6}\frac{\sum_{i=1}^{3}(\lambda_{i}-\bar{\lambda})^{2}}{\bar{\lambda}^{2}}.
\end{equation}
By definition, an inequality holds: $0\leq A\leq 1$ \cite{Blavatska2011}. The asphericity equals zero
for an ideally spherical shape, when all eigenvalues are equal: $\lambda_i=\bar{\lambda}$.
On the other extreme, an asphericity of a rod, when all eigenvalues but one vanish, $A=1$.
In the course of simulations, the instantaneous value $A$ is then
averaged over time trajectory of simulations, this averaged value is
denoted as  $\langle A\rangle$.
%

\section{Results}\label{III}

As already discussed in Sec.~\ref{II}, we consider four types of
heterogeneous stars and a homogeneous star copolymer, see
Fig.~\ref{Stars}. The linear size of a cubic simulation box is
chosen to be at least four gyration radius of a single branch in a
coiled conformation. For each heterogeneous star we perform $2
\cdot 10^6$ DPD steps, whereas for a homogeneous star the simulation
length is $6 \cdot 10^6$ DPD steps, for each considered value of $a_{CS}$.
The time step is $\bigtriangleup t^* = 0.04$. To
estimate statistical errors of particular characteristic $x$, the
simulation time trajectory is divided into four equal intervals. In
each i$th$ interval we calculate its partial simple average value
$\langle x\rangle^{[i]}$ and the histogram for its distribution
$P^{[i]}(x)$.
\begin{figure}[!h]
\centering
\includegraphics[width=8cm,angle=270]{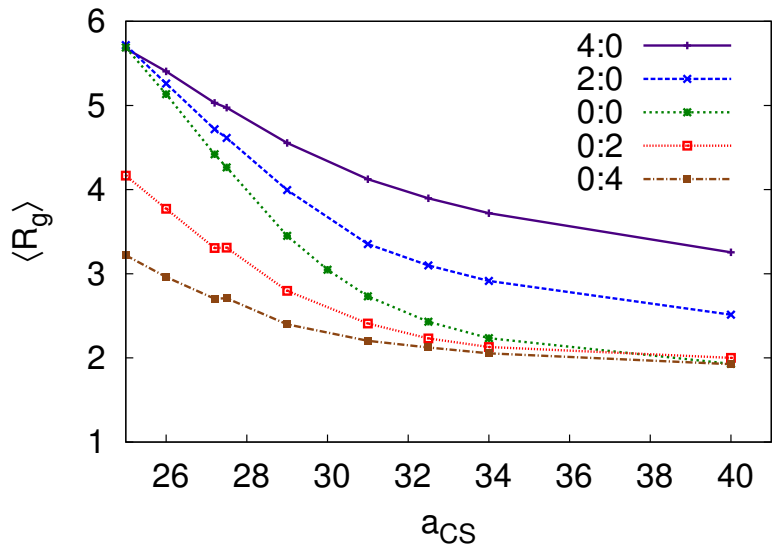}\hspace{-3em}
\caption{\label{fig1} Mean gyration radius $\langle R_g\rangle$ of
the ($f_A:f_B$) polymer star  as a function of
$a_{CS}$. Here $f_A$ is a number of solvophilic branches and $f_B$
is a number of solvophobic branches. The rest $f_C=f-f_A-f_B$ branches
have variable solvophilicity. The $a_{CS}$ parameter discriminates
between good ($a_{CS}=25$) and bad ($a_{CS}=40$) solvent quality for the $f_C$ branches.
Green line corresponds to a homogeneous star when $f_C=f$, marked as ($0:0$). Here the gyration radius is expressed in the units $\sigma$ which is the diameter of one monomer.}
\end{figure}

 First we analyse the value of the gyration radius $R_g$ for a star polymer
depending on the solvent quality. The latter was tuned by the choice
of $a_{CS}$, from the good solvent case ($a_{CS}=25$) to the bad one
($a_{CS}=40$), see Fig.~\ref{fig1}. Let us consider the change
undergone by $R_g$ for the case of the homogeneous star (0:0), when
$a_{CS}$ increases from $25$ to $40$. This change drives the
conformation of all the branches from a coiled state to the
collapsed one. As a consequence, the $R_g$ decreases smoothly from $5.8$
down to $2$. For the heterogeneous stars with fixed solvophilic
branches ($4:0$ and $2:0$), the $R_g$ value at $a_{CS}=25$ is,
obviously, the same as the one for the homogeneous star. But at
$a_{CS}=40$ the gyration radius for both $(4:0)$ and $(2:0)$ stars
have higher magnitude due to the fact that $f_A$ branches are still
in a coiled state, whereas the other $f_C$ branches are in a
collapsed state. On contrary, heterogeneous stars with fixed
solvophobic branches $(0:2)$ and $(0:4)$ have the same value of
$R_g$ at $a_{CS}=40$ as the homogeneous star and a lower value at
$a_{CS}=25$. This is, obviously, due to the fact that at $a_{CS}=25$
only $f_C$ branches are in a coil state, whereas the other $f_B$
branches are always in a collapsed state.

\begin{figure}[!h]
\centering
\includegraphics[width=8cm,angle=270]{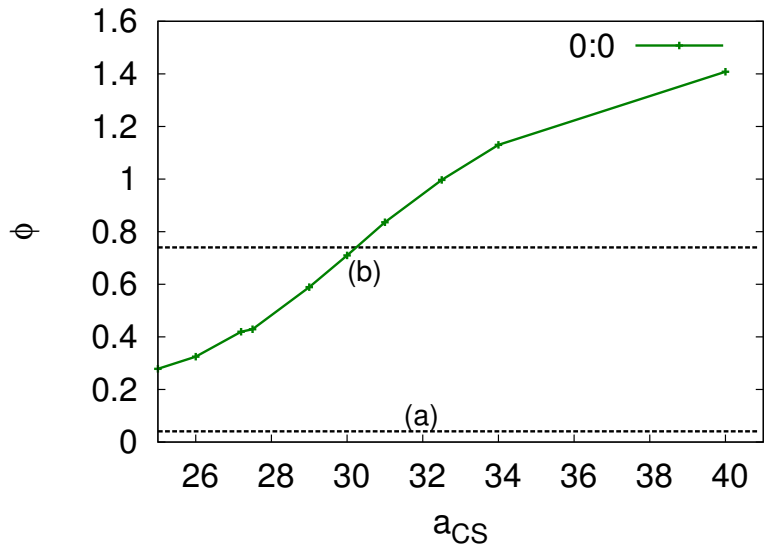}\hspace{-3em}
\caption{\label{fig8} Solid line: packing fraction $\phi$ of  the homogeneous star ($0:0$) at various $a_{CS}$, the solvophobicity of its beads. Dashed lines: packing fractions of the case when all branches are fully stretched (a) and when they are tightly packed(b)}
\end{figure}

In the DPD method all the beads are soft and can overlap,
unlike for the hard spheres model. To quantify this effect,
we analysed the packing fraction $\phi$ of beads depending on the
solvent quality, see Fig.~$\ref{fig8}$. It is given by
\begin{equation}\label{frac}
\phi = \frac{N_b \frac{4}{3}\pi r^3}{V_{glob}}
\end{equation}
where $N_b$ is the number of beads in a star polymer, $r$ is the
radius of a single bead and $V_{glob}$ is the volume occupied by star
polymer. This volume can be expressed $via$ the expression $V_{glob}
= \frac{4}{3} \pi R^3$, where $R$ is the radius of an equivalent
sphere with the same volume as that of a star polymer. We estimate $R$ from
its relation to the gyration radius $R_g$. As the first approximation,
one can take the relation $R=\sqrt{\frac{5}{3}}R_{g}$ which holds for a
solid spheres. We also display in Fig.~$\ref{fig8}$ the packing
fractions that relate to two limiting cases. These are marked as
$(a)$: the case when all the branches are in a fully stretched conformation and
$(b)$, when the star polymer comprises a compact object of tightly
packed hard spheres. In the good solvent regime at $a_{CS}=25$, the
packing fraction is $\phi\approx 0.28$ which is close to the case
$(a)$. At $a_{CS}=30$, the packing fraction crosses the line $(b)$.
With an increase of $a_{CS}$ the packing fraction increases, and in
the bad solvent regime at $a_{CS}=40$, $\phi$ is equal to $1.4$. It
is indicating severe beads overlap, and, hence highly compressed state of the star polymer.

\begin{figure}[!h]
\centering
\includegraphics[width=8cm,angle=270]{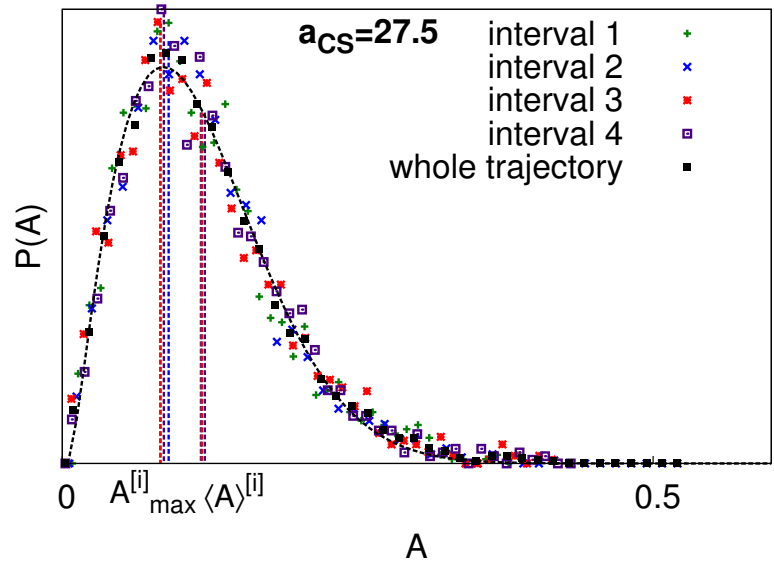}\hspace{-3em}
\caption{\label{fig5}Distribution of the asphericity $A$ of a homogeneous star at solvent
quality $a_{CS}=27.5$ for the intervals 1: $0<N_{stp}<15\cdot10^5$,
2: $15\cdot10^5<N_{stp}<3\cdot10^6$, 3:
$3\cdot10^6<N_{stp}<45\cdot10^5$, 4:
$45\cdot10^5<N_{stp}<6\cdot10^6$, as well as for the whole simulation
$0<N_{stp}<6\cdot10^6$. Black curve is fitting according to Eq.
\ref{Lhuillier} }
\end{figure}

\begin{figure}[!h]
\centering
\includegraphics[width=8cm,angle=270]{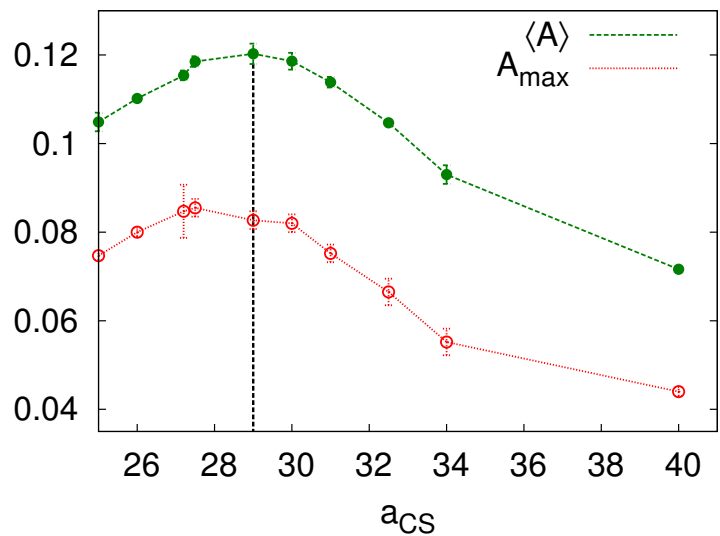}\hspace{-3em}
\caption{\label{fig6} Plot of the asphericity of a homogeneous star as a function of the
solvophobicity of its beads $a_{CS}$ . Here $\langle
A\rangle$ and $A_{max}$ are the mean and maximal values of $A$ is found from the distribution (\ref{Lhuillier}).}
\end{figure}

    Let us switch our attention to the asphericity of the stars.
The histograms $P^{[i]}(A)$  for the distribution of asphericity in each $[i]$th interval $A$ are presented in Fig.~\ref{fig5} for the case of homogeneous star at
$a_{CS}=27.5$. This value was estimated earlier in
Ref.~\cite{Nardai2009} as the one providing approximately the
$\theta$-conditions for similar model of a star polymer. We observe
that the data points obtained for each $P^{[i]}(A)$ follow the same
curve. The shape of this curve, similarly to the case of linear
polymer chain \cite{Kalyuzhnyi2016}, can be approximated well by the
Lhuillier-type distribution:
\begin{equation}\label{Lhuillier}
P_L(A) = B \exp \left[-\Big(\frac{A'}{A}\Big)^{\epsilon_1} -
\Big(\frac{A}{A'}\Big)^{\epsilon_2}\right],\;\;\;
A_{max}=A'\Big(\frac{\epsilon_1}{\epsilon_2}\Big)^{\frac{1}{\epsilon_1+\epsilon_2}}.
\end{equation}
Here $A_{max}$ is the position of the maximum for the $P_L(A)$ and
$B,A',\epsilon_1$ and $\epsilon_2$ are fitting parameters. We
perform the separate fit for each $P^{[i]}(A)$ to the form Eq.~(\ref{Lhuillier}) resulting in a set of
parameters $B^{[i]},A'^{[i]},\epsilon_1^{[i]},\epsilon_2^{[i]}$ and
$A_{max}^{[i]}$. The sets provide both the respective averages for
each of these parameters, as well as their standard deviations evaluated within the set. The
same procedure is used for all values of $a_{CS}=25-40$. The results
for the simple average of the asphericity $\langle A\rangle$, and
the average position of the maximum $A_{max}$, alongside with their
respective errors are shown in Fig.~\ref{fig6}. Similarly to the
data shown in Fig.~\ref{fig5}, the magnitude of $\langle A\rangle$
is higher than that for $A_{max}$. However, the changes undergone by
both characteristics  with the variation of $a_{CS}$ are very
similar. Namely, both increase within the interval $25<a_{CS}<29$,
peak at approximately $a_{CS}=29$ and then both decay. For both
characteristics we observe a maximum at the interval $27<a_{CS}<29$.
This interval contains the value $a_{CS}=27.5$ estimated as a
$\theta$-point by Nardai and Zifferer \cite{Nardai2009}, for the
same simulations model. Below we attempt to clarify the reason for
this maxima.

\begin{figure}[!h]
\centering
\includegraphics[width=8cm,angle=270]{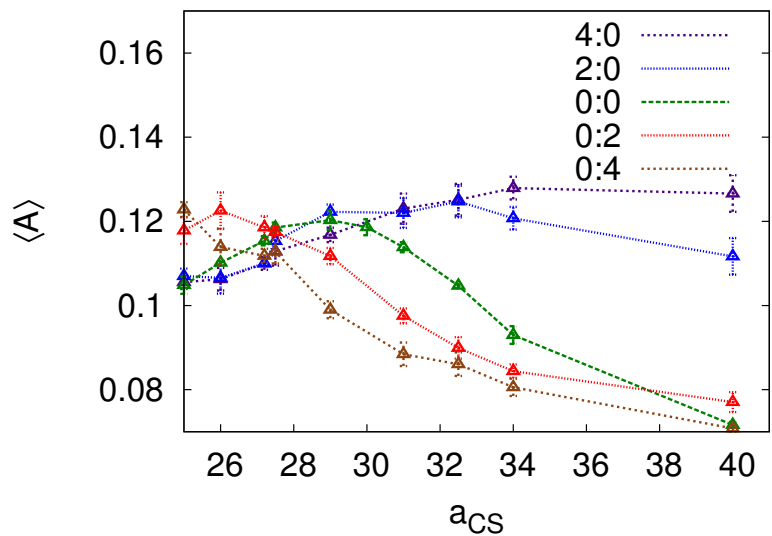}\hspace{-3em}
\caption{\label{fig3}Asphericity $A$ of the ($f_A:f_B$) polymer star
as a function of $a_{CS}$, the solvophobicity of the $C$ beads Here $f_A$ is a number
of solvophilic branches and $f_B$ is a number of solvophopic
branches . Factor $a_{CS}$ discriminates between good ($a_{CS}=25$)
and bad ($a_{CS}=40$) solvent quality. Green line corresponds to a
homogeneous star with all branches of changing solubility ($0:0$).}
\end{figure}

  To clarify this peculiarity of the asphericity at the $\theta$-condition,
we will consider the asphericity of the heterogeneous stars,
depicted in Fig.~\ref{fig3}. The cases $(2:0)$ and $(4:0)$ at
$a_{CS}=25$ are characterized by all the branches being in a coiled
state, similarly to the case $(0:0)$ at the same $a_{CS}$. Hence,
the same value of $\langle A\rangle = 0.105$ is observed in all
cases mentioned above. With the $a_{CS}$ value approaching 40, the $(0:2)$ and
$(0:4)$ stars at $a_{CS}=40$ have all their branches collapsed.
Again, this mimics the case of $(0:0)$ star at $a_{CS}=40$.
Therefore, not surprisingly, the $\langle A\rangle$ values for those
stars are close to the respective value of  $0.075$ for the
homogeneous star (cf. Figs.~$\ref{fig6}$ and $\ref{fig3}$). However,
cases $(0:2)$ and $(0:4)$ at $a_{CS}=25$ both have a higher
asphericity value of $\langle A\rangle =0.12$, the same as for the
case $(2:0)$ and $(4:0)$ at $a_{CS}=40$. This higher value of
$\langle A\rangle$ is observed due to the fact that in all these
cases some branches are in a coiled state and the other branches are
collapsed. It is instructive to note that the same value $\langle A\rangle
=0.12$ is also observed for the homogeneous star ($0:0$) in the
interval $27<a_{CS}<29$. This leads us to the idea that the maxima
of $\langle A\rangle$ for the ($0:0$) star, observed in this
interval might be related to the coexistence of both coiled and
collapsed configurations of the branches, in a $\theta$-solvent
regime.

\begin{figure}[!h]
\centerline{\includegraphics[width=0.25\textwidth,angle=270]{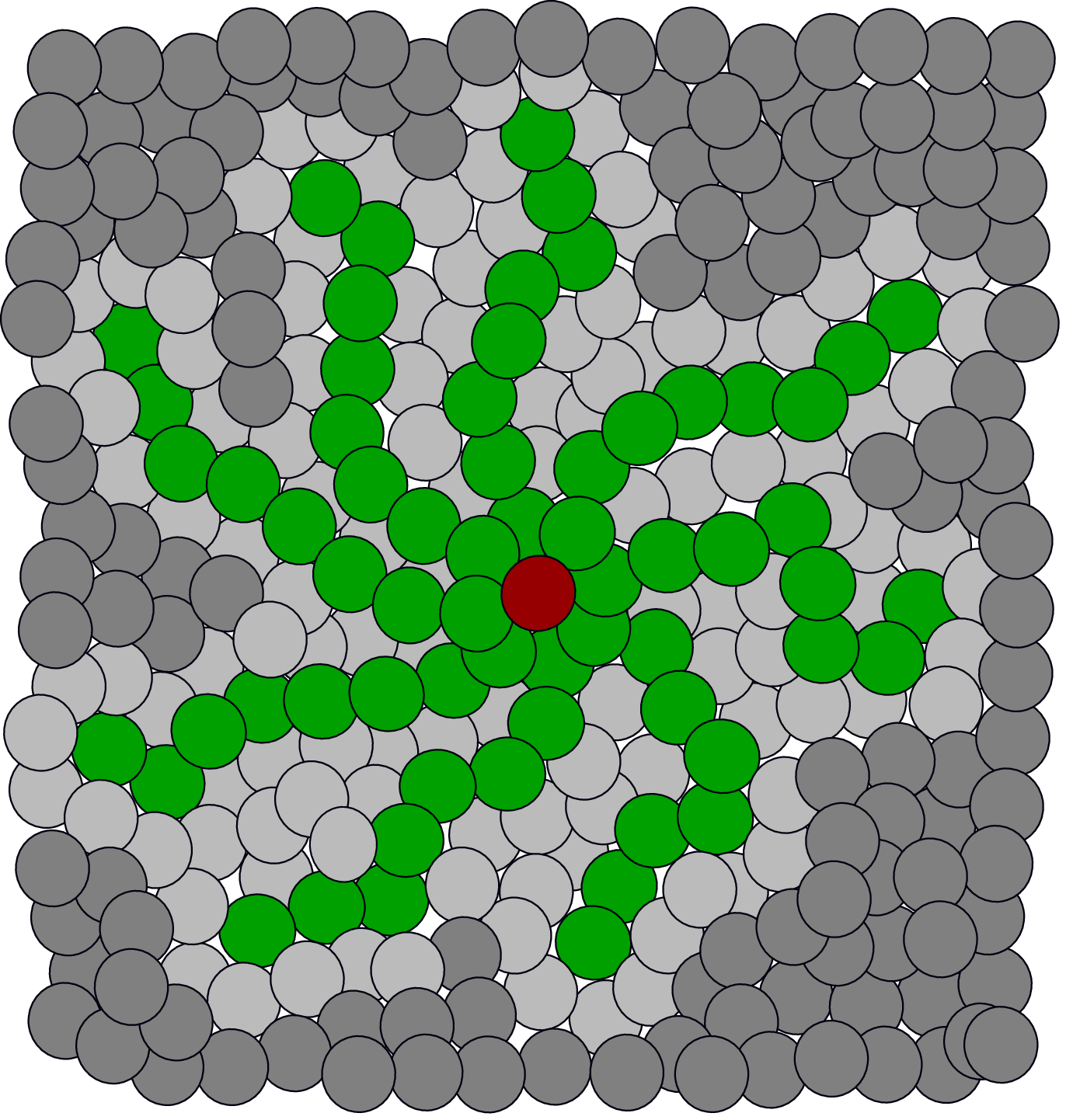}\hspace{3em}
\includegraphics[width=0.25\textwidth,angle=270]{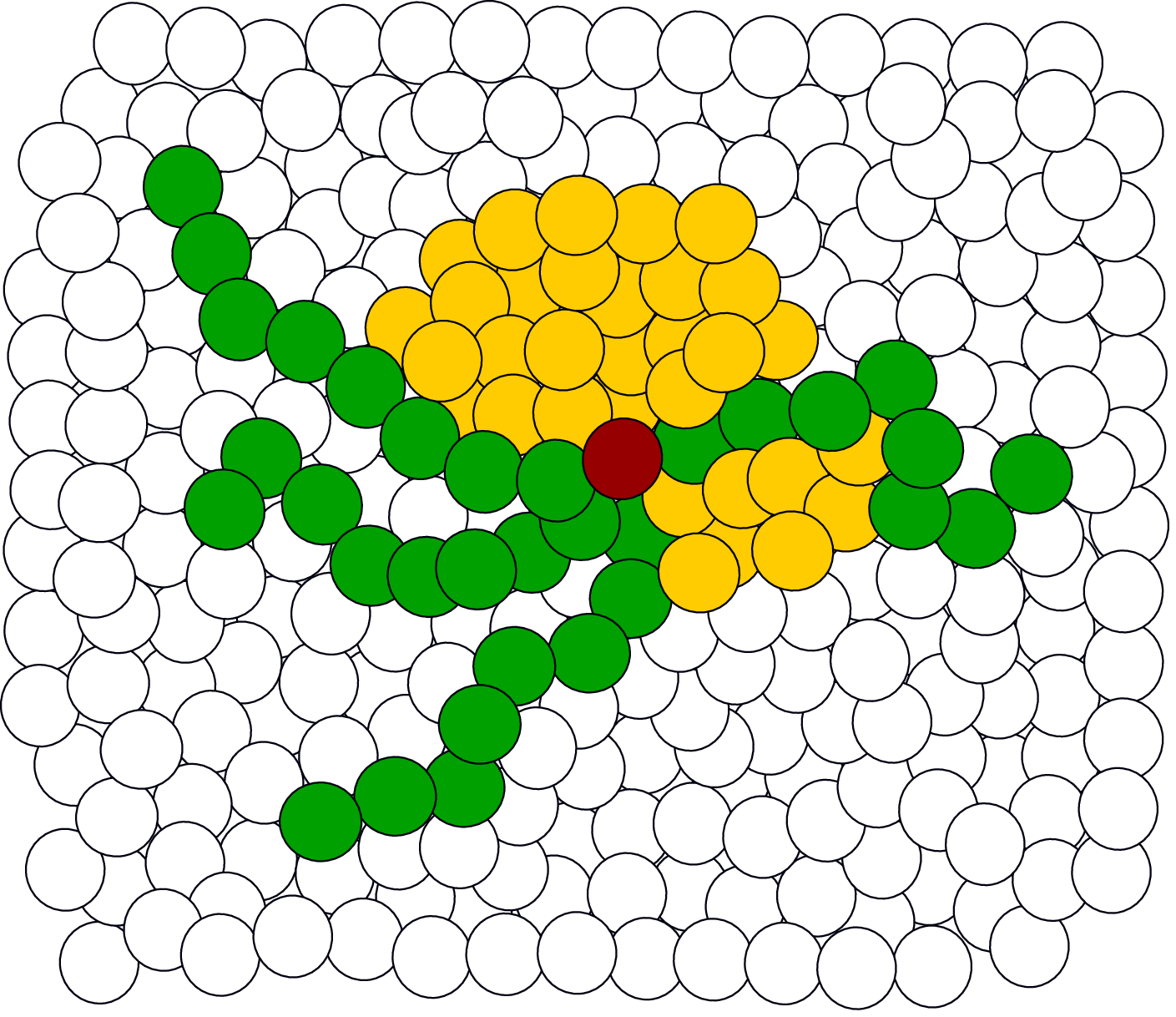}\hspace{3em}
\includegraphics[width=0.25\textwidth,angle=270]{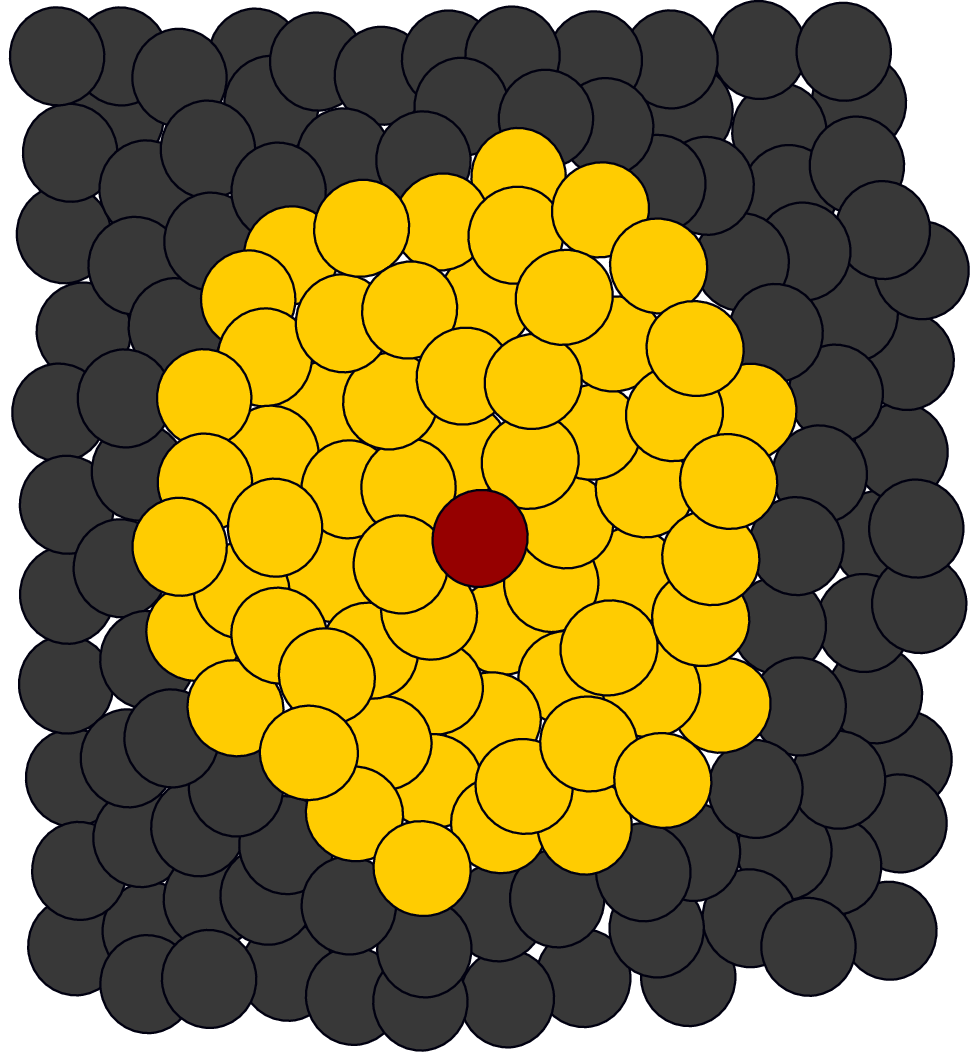}}
\centerline{$(a)$ \hspace{10em}  $(b)$ \hspace{12em} $(c)$}
\caption{\label{fig9} Schematic illustration for the coformation of
individual branches of homogeneous star polymer $(0:0)$ in various
regimes. $(a)$ good solvent, $(b)$ $\theta$-condition and $(c)$ bad
solvent.}
\end{figure}

There different conditions for the individual branches of
homogeneous star polymer, namely the cases of good, bad and
 the $\theta$-solvent, are illustrated in Fig.~$\ref{fig9}$.
These cases are considered in terms of the interplay between the
enthalpic, $U$, and the entropic $S$, contributions to the free
energy $F=U-TS$, with $T$ being the temperature. The case of a good
solvent ($a_{CS}=25$) is displayed in Fig.~$\ref{fig9}(a)$. Here all
the branches are in a coiled state and are surrounded by a solvation
shell. In this case, there is an effective repulsion between
branches, leading to the non-zero enthalpy contribution $U$. For the
case of a bad solvent, Fig.~$\ref{fig9}(c)$, the branches strongly repel each other
strongly by a solvent, leading to their collapse, this is
interpreted as the effective solvent-mediated attraction between
branches. In this case, the contribution of $U$ to the free
energy is also non-zero. The intermediate case, where polymer is in the
$\theta$-solvent condition, is characterized by vanishing of the
enthalpy contribution $U$ and the free energy is driven now only by
the entropy. As a consequence, all branches turn into the ideal
chains, their conformations are uncorrelated and each one can be interpreted
as a random walk. This may lead to the coexistence of both coiled
and collapsed configurations of individual chains, as illustrated schematically in
Fig.~$\ref{fig9}(b)$.

\begin{figure}[!h]
\centering
\includegraphics[width=8cm,angle=270]{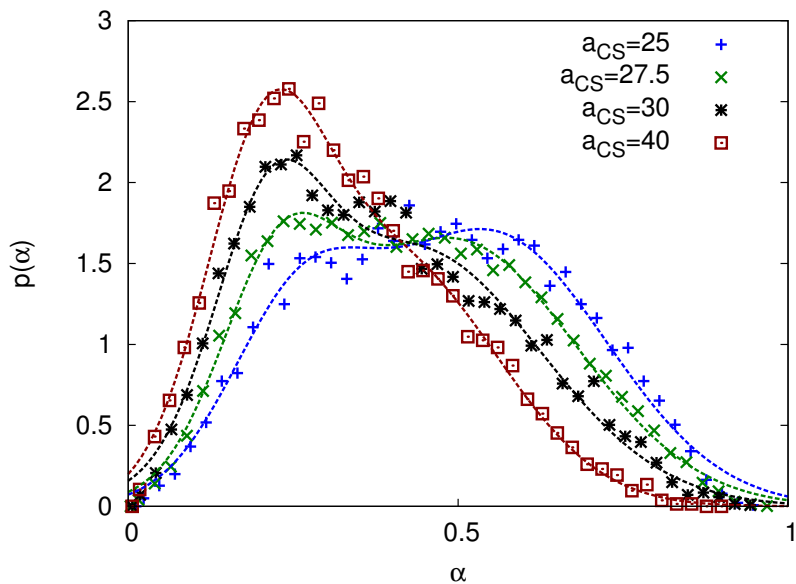}\hspace{-3em}
\caption{\label{fig7}Distribution of asphericity $p(\alpha)$ of individual
branches of a homogeneous star at various solvophobicity of its branches $a_{CS}$.}
\end{figure}

To confirm this interpretation of the behavior of the star polymer
in the $\theta$-solvent we conducted the analyses of the
configurations of individual branches. The histogram for the
distribution $p(\alpha)$ for the asphericity $\alpha$ of individual branches
is illustrated in Fig.~$\ref{fig7}$ for the homogeneous star at
$a_{CS}=25,27.5,30$ and $40$. The general shape of this distribution
changes essentially when $a_{AC}$ varies from $25$ to $40$. In
general, it shows the presence of two maxima. Therefore, we fitted
the data by a double Gaussian distribution:

\begin{equation}\label{Gausse}
p(a) = A' \exp \left[-\Big(\frac{{x-\alpha'}}{\sigma'}\Big)^{2}\right] +
A'' \exp \left[-\Big(\frac{{x-\alpha''}}{\sigma''}\Big)^{2}\right],
\end{equation}
where $A'$ and $A''$ are respective weights, $\alpha'$ and $\alpha''$ -
respective position of two maxima, $\sigma_1$ and $\sigma_2$ are the
respective dispersions. The results of the fits for $A',A'', \alpha'$
and $\alpha''$ are collected in Tab. $\ref{tab_aver}$.
\begin{table}
\begin{center}
\begin{tabular}{|l|l|l|l|l|}
  \hline
  $a_{CS}$ & 25 & 27.5 & 30 & 40 \\ \hline
  $\alpha'$ & 0.257 & 0.225 & 0.212 & 0.205 \\ \hline
  $A'$ & 1.056 & 1.266 & 1.289 & 1.811 \\ \hline
  $\alpha''$ & 0.555 & 0.512 & 0.435 & 0.402 \\ \hline
  $A''$ & 1.683 & 1.617 & 1.587 & 1.514 \\
  \hline
\end{tabular}
\caption{\label{tab_aver} $\alpha'$ and $\alpha''$ are the positions of the
maxima, $A'$ and $A''$ are the respective weights of the Gaussian
distributions and $\sigma'$ and $\sigma''$ are the respective
dispersion.}
\end{center}
\end{table}
 Based on these results and the shape of the $p(a)$ distribution, we can
deduct the following conclusion. In all cases $a_{CS}=25-40$ we see
the coexistence of typical conformations, one with lower asphericity,
$\alpha'\approx 0.2-0.26$, and another one with a higher asphericity,
$\alpha''\approx 0.4-0.56$. For the case of $a_{CS}=40$ low asphercity
conformations prevail, as far as $A'>A''$. Here all branches are in
the collapsed state and the distribution has one maximum near
$\alpha'=0.25$. On the contrary, at $a_{CS}=25$ we have $A'<A''$, hence
the conformations with higher aspericity prevail. This is due to the
branches being predominately found in a coiled state. At $a_{CS}=27.5-30$ the
distribution clearly has two maxima. This evidences the coexistence
of rather collapsed conformations with those that are more close to
a coiled state. Hence, at the $\theta$ condition of a star polymer, there is much higher degree of the solvent regime for conformational freedom of its
branches, and as a consequence, by a
coexistence of more coiled conformation with the ones closer to a
collapsed state. This explains the maximum for the average asphericity $\langle A\rangle$
as observed in Fig.~\ref{fig6}.

\section{Conclusions}\label{IV}
We have analysed the shape characteristics of the coarse-grained
heterogeneous and homogeneous star-like polymers at various solvent
quality using the DPD simulations. The heterogeneous star is
characterized by different solvophobicity of its individual
branches. The solvent quality was tuned by varying the parameter
$a_{CS}$ for the repulsive conservative force of the branches with variable solvophobicity. For homogeneous and
four types of heterogeneous star polymers, the gyration radius
decreases with the increase of the repulsion parameter $a_{CS}$
indicating a collapsed state for the part or all their branches. The packing
fraction $\phi$ at $a_{CS}=30$ is close to the packing
fraction of the hexagonally packed hard spheres. We
found an interesting effect that, upon the change of solvent
properties, the asphericity of a homogeneous star reaches its
maximum value when the solvent is near $\theta$-point.
The effect is explained by the interplay between the enthalpic and
entropic contributions to the free energy. In particular, at the
$\theta$-point condition, the enthalpic contribution vanishes and
the branch conformation are driven exclusively by the entropy.
This provides means for the wider spectra of possible conformations
from collapsed to the coiled ones. We check this explanation by the
analyses of the asphericity of the individual branches. Their
distribution  near $\theta$-condition has two maxima, which is well fitted by double
Gaussian distribution. This  confirms
the coexistence of two types of conformation, namely, more coiled and more collapsed
ones. An extension of this  analyses to the case of aggregation of
star-like polymers into micelles which has applications for the drug
delivery systems will be the subject of the forthcoming paper.

\section{Acknowledgements}\label{V}
This work was supported in part by FP7 EU IRSES project No. 612707 "Dynamics of and in Complex Systems"
and No. 612669 "Structure and Evolution of Complex Systems with Applications in Physics and Life Science". J.I. acknowledges Alexander Blumen for stimulating discussions towards the interpretation of the result.

\end{document}